\documentclass[12pt]{article}%
\usepackage{amsmath}
\usepackage{graphicx}
\usepackage{amsfonts}
\usepackage{amssymb}
\usepackage{color}%
\newtheorem{theorem}{Theorem}

\newtheorem{corollary}{Corollary}

\usepackage{graphicx}

\newtheorem{remark}{Remark}

\newenvironment{proof}[1][Proof: ]{\textbf{#1}\it }

\begin{document}
\title{On trend and its derivatives estimation in repeated time series 
with subordinated long-range dependent errors}
\author{Haiyan Liu and Jeanine Houwing-Duistermaat\\Department of Statistics\\University of Leeds}
\maketitle

\begin{abstract}
For temporal regularly spaced datasets, a lot of methods are available and the properties
of these methods are extensively investigated. 
Less research has been performed on irregular temporal datasets subject to measurement error
with complex dependence structures, while this type of datasets is widely available.
In this paper, the performance of kernel smoother for trend and its derivatives
is considered under dependent measurement errors and irregularly spaced sampling scheme.
The error processes are assumed to be subordinated Gaussian long memory 
processes and have varying marginal distributions.
The functional central limit theorem for the estimators of trend and its derivatives 
are derived and bandwidth selection problem is addressed.
\end{abstract}


\section{Introduction}
In functional data analysis (FDA) studies, it is assumed that one observes $n$ independent
random curves  $X_i(t)=\mu(t)+\sum_{l=1}^\infty\xi_{il}\phi_l(t)$ ($i=1, ..., n$), which
come from an underlying random process $X(t)\in L^2[0, 1]$.
Here $\mu(t)$ is the population mean, the coefficients $\xi_{il}$ are uncorrelated random variables 
with mean zero and variance $\lambda_l$ ($\sum \lambda_l<\infty$), 
and the functions $\phi_l$ build an orthonormal $L^2[0, 1]$ basis.
Note that $\lambda_l$ and $\phi_l$ are eigenvalues and eigenfunctions of the covariance operator
$C(y)=\mathbf E[\langle X, y\rangle X],\ y\in L^2[0, 1]$, respectively.
For general overview on FDA see e.g. Ramsay and Silverman(2005), Horv\'{a}th and Kokoszka (2012)
and references therein.

In practice, each random curve is typically observed at discrete but not
necessarily equidistant time points $t_{ij}\in[0, 1]$ ($j=1, ..., N_i$) 
and observations are perturbed by random noise.
That is, we observe error perturbed curves
$$Y_{ij}=\mu(t_{ij})+\sum_{l=1}^\infty\xi_{il}\phi_l(t_{ij})+\epsilon_{ij}$$
where $\epsilon_{ij}$ are mean zero random measurement errors. 
Existing work assumes that 
the random errors $\epsilon_{ij}$ are independent and Gaussian, 
see Cai and Yuan (2010), Peng and Paul (2009), Staniswalis and Lee (1998), 
and Yao et al. (2005).

However, the assumption of independence is too restrictive. 
For example, for certain types of 
EEG observations the dependence or even long-range dependence in the error processes
$\epsilon_{ij}$ ($j\in\mathbb N$) occurs (see e.g. Bornas et al. 2013, Linkenkaer-Hansen et al. 2001,
Nikulin and Brismar 2005, Watters 2000). 
Beran and Liu (2014, 2016) therefore consider the 
estimation of trend and covariance for dense functional data
that are perturbed by long-range dependent Gaussian errors.
For details on statistical inference for long-range dependent processes,
see e.g. Beran (1994), Giraitis et al. (2012), and Beran et al. (2013)
and references therein.

In addition, in many applications, the Gaussianity and equidistance assumption is unrealistic. 
For example, in observational studies in medicine, the patient visits reported
in electronic health records are irregular and sparse.
Moreover, the presence of dependent non-Gaussian measurement errors
makes the analysis challenging.
Another example, in the context of ecology and climate science, is plaeo-proxy irregular time series 
perturbed by non-Gaussian errors as has been reported (see e.g. Men\'endez et al. 2010).
One more example, in the context of neurobiology, is the phenomenon 
that subodinated long-range denpendent errors 
exist in the optical measurements of the calcium concentration in a glomerulus of the antennal
lobe of a honey bee after an olfactory stimulus as reported in Beran and Weiershauser (2010).

For single time series, Men\'endez et al. (2010) and Men\'endez et al. (2013)
investigate the trend and its derivatives 
kernel estimation for observations perturbed by subordinated Gaussian long memory errors.
Ghosh, S. (2014) addresses estimation of trend and slope functions in a partial linear model
when the errors are unknown time-dependent functionals of latent Gaussian processes.
Beran and Weiershauser (2011) consider estimation in a spline regression
model with long-range dependent errors that are generated by Gaussian subordination.

The main purpose of this paper is, for repeated time series,
to investigate the influence of subodinated
long-range dependent Gaussian errors on the trend and its derivatives estimation.
Detailed results on the following-up analysis on covariance and princpial components
will be reported in a subsequent paper.

The rest of this paper is organised as follows.
Section 2 explains the model in details and gives the kernel estimation of
the global mean and its derivatives.
In Section 3, we investigate the asymptotic bias and variance 
and establish the functional central limit theorem of our estimator.
The proofs of the theorems appear in Appendix.

\section{Definitions}
\subsection{The model}

We consider the square-integrable random function $X(t)$ defined over $[0, 1]$
with mean $\mu(t)=E[X(t)]$ and continuous covariance $C(s, t)=Cov(X(s), X(t))$.
Then, by Karhunen-Lo\`eve (K.L.) expansion, $X(t)$ has the form
\begin{equation*}
X(t)=\mu(t)+\sum_{l=1}^{\infty}\xi_{l}(\omega)\phi_l(t)
\quad (
t\in[0, 1])
\end{equation*}
where $\{\xi_l\}$ are pairwise uncorrelated random variables 
and functions $\{\phi_l\}$ are continuous real-valued functions on $[0, 1]$ 
that are pairwise orthogonal in $L^2$.

Let $X_i$, $i=1,...,n$, be unobservable i.i.d. copies of $X$.
For the $i$th subject, measurements are available at time points $t_{ij}$, $j=1,...,N_i$, 
and the observations at these time points are perturbed by 
subordinated long-range dependent noise,
so the actual observations $Y_{ij}$ are:
\begin{equation}
\label{Equation Model Observation}
Y_{ij}=X_i(t_{ij})+\epsilon_i(j)
\quad (i=1,...,n; j=1,...,N_i)
\end{equation}
where $n$ is the number of subjects,
$N_i$ is the number of sampling points for $i$-th subject,
$T_{ij}\in \mathbb N_+$, $T_{i1}\leq ... \leq T_{iN_i}\leq T_{max}=\underset{i}\max\{T_{iN_i}\}$, 
and $t_{ij}=T_{ij}/T_{max}\in[0, 1]$.
Specifically,
\begin{itemize}
\item By K.L. expansion, the random curves $X_i(t)$ follow:
\begin{equation}
\label{Equation Model Curve}
X_i(t)=\mu(t)+\sum_{l=1}^{\infty}\xi_{il}\phi_l(t)
\quad (t\in[0, 1])
\end{equation}
\item 
Let $Z(u)$ ($u\in\mathbb R$) be a continuous time stationary Gaussian process
with $E[Z]=0$, $var(Z)=1$ and
\begin{equation*}
\label{Equation model subordination error autocovariance}
\gamma_Z(v)
=cov\left(Z(u),Z(u+v)\right)
\underset{v\to\infty}{\sim}c_Zv^{2d-1}
\end{equation*}
where $0<d<\frac12$, ``$\sim$''\ means that the ratio of the left and right hand side tends to one.
For each sample $i$, the error process $\{\epsilon_i(j)\}$ ($j\in\mathbb{N}$)
has the form
\begin{equation}
\label{Equation model error}
\epsilon_i(j)=G(Z(T_{ij}), t_{ij}).
\end{equation}
For each fixed $t\in[0, 1]$ the function $G(\cdot, t)$ is assumed to be in $L^2$-space
of functions (on $\mathbb R$) with $E[(Z, t)]=0$
and $\|G\|^2=E[G^2(Z, t)]<\infty$.
This implies a convergent $L^2$-expansion

\begin{equation*}
\label{Equation model error autocovariance}
G(Z, t_{ij})=\sum_{k=q}^\infty \frac{c_k(t_{ij})}{k!}H_k(Z)
\end{equation*}
where $H_k(\cdot)$ are Hermite polynomials, $q\geq1$ is the so-called Hermite rank,
and $c_k(t)=E[G(Z, t)H_k(Z)]$.

\item $\xi$ and $\epsilon$ are independent of each other.
\end{itemize}

\begin{remark}
The function $G$ provides the possibility of having non-Gaussian errors with a changing
marginal distribution.
First, note that when $G$ is nonlinear, $\epsilon$ is non-Gaussian.
Therefore, $G$ can represent the departure from the Gaussian assumption
which provides more flexibility for modelling.
Second, a proper choice of $G$ can match any marginal distibution for $\epsilon$,
which depends on $c_k(t)$.
Third, from the view of robustness, $\epsilon$ can be seen as a perturbed version
of $Z$ and if $\epsilon$ is close to $Z$ then $G$ is close to the identity function.

\end{remark}

\subsection{The estimation}
To estimate $\mu^{(v)}(t)$, since each time series is unevenly recorded, 
we consider a Priestley-Chao type kernel estimator defined by
\begin{equation}
\label{Equation trend estimation - PC}
\hat\mu^{(v)}(t)
=\frac{(-1)^v}{b^{v+1}} \frac{1}{n}\sum_{i=1}^n\sum_{j=1}^{N_i}(t_{ij}-t_{i,j-1}) 
  K^{(v)}\left(\frac{t_{ij}-t}{b}\right)Y_{ij}
\end{equation}
where $t_{i0}=0$ and $K$ is a kernel function satisfing the following general assumptions
(Gasser et al 1985, Men\'endez, Ghosh and Beran 2010):
\begin{itemize}
\item (K1) $K\in C^{v+1}[-1, 1]$.
\item (K2) $K(x)\geq 0, \ K(x)=0 \ (|x|>1), \ \int K(x)dx=1$.
\item (K3) $K^{(v)}(\cdot)$ is Lipschitz continuous,
i.e. there exists $L\in\mathbb R_+$ such that,
for any $x, y\in[-1, 1]$,
$$|K^{(v)}(x)-K^{(v)}(y)|\leq L|x-y|.$$
\item (K4) $K(\cdot)$ is of order ($v, k$), $v\leq k-2$, where $k\in\mathbb N_+$, i.e.,
\begin{equation*}
\int_{-1}^1 x^jK^{(v)}(x)dx=\left\{
   \begin{aligned}
   (-1)^v v!, \quad & j=v  \\
   0, \quad & j=0, ..., v-1, v+1, ..., k-1 \\
   \theta, \quad & j=k\\
   \end{aligned}
   \right.
\end{equation*}
where $\theta\ne0$ is a constant.
\item (K5) $K^{(j)}(-1)=K^{(j)}(1)=0$ for all $j=0,1,..., v-1$.
\item (K6) $0\leq \kappa_{v+1}=\sup_{x\in[-1, 1]}|K^{(v+1)}(x)|<\infty$.
\end{itemize}

\section{Asymptotic results}
\subsection{Assumptions}
Apart from the general assumptions on kernel function, the following technical
conditions on the time points are required:
\begin{itemize}
\item (T1) $c_k(t)$ are continuously differentiable w.r.t. $t\in[0, 1]$.
\item (T2) $\frac12-\frac1{2q}<d<\frac12$.
\item (T3) For $i=1,...,n$,
$$0\leq T_{i1}\leq...\leq T_{iN_i}\le T_{max}=\underset{i}\max\{T_{iN_i}\}, 
\ t_{ij}=T_{ij}/T_{max}\in[0, 1].$$
\item (T4) For $i=1,...,n$,
$$\alpha_{N}^{-1}\leq \underset{i}\min\{t_{ij}-t_{i,j-1}\}
\leq \underset{i}\max\{t_{ij}-t_{i,j-1}\}\leq \beta_{N}^{-1}$$
where $\alpha_{N}\geq\beta_{N}>0$ and $\beta_{N} \to \infty$ 
as $N=\underset{i}\min\{N_i\}\to\infty$.
% \item (T5) $n \to \infty, b\to0, \ bT_N\to \infty$, and $b\beta_{N}\to \infty$.
\item (T5) $\underset{{N}\to\infty}\lim(b\alpha_{N})^{1+(1-2d)q}(b\beta_{N})^{-2}=0$.
\end{itemize}

\begin{remark}
Assumption (T1) implies that the residuals have a slowly changing marginal distribution.
Further, (T1) also includes the necessary conditions for the weak convergence of $\mu^{(v)}(t)$.
Since $cov(Z(T_{ij1}),Z(T_{ij_2}))\sim |T_{ij1}-T_{ij_2}|^{2d-1})$ implies
$cov(\epsilon_{ij_1}, \epsilon_{ij_2})\sim  |T_{ij1}-T_{ij_2}|^{(2d-1)q})$,
the condition of (T2) implies that the long-range dependence of $Z$ 
is inherited by the subordinated process $\epsilon$.
Assumption (T5) is a condition needed to obtain an asymptotic approximation of 
the variance of $\mu^{(v)}(t)$.
\end{remark}

The trend $\mu(t)$ and the basis $\phi_l(t)$ are assumed to have the following 
smoothness:
\begin{itemize}
\item (M1) $\mu\in C^{k}[0, 1]$.
\item (M2) $\phi_l\in C^{k}[0, 1]$.
\end{itemize}

\subsection{Bias and variance of $\hat\mu^{(v)}$}

\begin{theorem}
\label{Theorem - Bias and Var of mean and its der}
Let $Y_{ij}$ be defined by (\ref{Equation Model Observation})-(\ref{Equation model error}).
Suppose the assumptions (K1)-(K6), (T1)-(T5), (M1) and (M2) hold. 
Let $n\to \infty$, $N\to\infty$, $b\to 0$, $bT_{max}\to\infty$.
Moreover, let 
\begin{align}
\label{Condition - b and beta_N}
b^{k+1} \beta_{N}\to\infty.
\end{align} 
The following holds for any $t\in(0, 1)$:
\begin{itemize}
\item Bias:
\begin{align} 
\label{bias}
&E\left[\hat\mu^{(v)}(t)\right]-\mu^{(v)}(t) \nonumber\\
&= b^{k-v}C_{bias,v,k}(t)+o(b^{k-v})+O\left((b^{v+1}\beta_N)^{-1}\right) \nonumber\\ 
&= b^{k-v}C_{bias,v,k}(t)\left(1+o(1)\right)
\end{align}
where
$$C_{bias, v,k}(t)=\frac{\mu^{(k)}(t)}{k!} \int K^{(v)}(x)x^{k}dx.$$
\item Variance:
\begin{align}
\label{variance}
&var\left(\hat\mu^{(v)}(t)\right) \nonumber\\
&=n^{-1}C_{var, v}(t)\left(1+O(b^{k-v})+O\left((b^{v+1}\beta_N)^{-1}\right)
     +O\left(b^{-2v}(T_{max}b)^{(2d-1)q}\right)\right) \nonumber\\
&=n^{-1}C_{var, v}(t)\left(1+O(b^{k-v})+O\left(b^{-2v}(T_{max}b)^{(2d-1)q}\right)\right)
\end{align}
where
$$C_{var, v}(t)=\sum_{l=1}^\infty\lambda_l\left(\phi_l^{(v)}(t)\right)^2.$$
\end{itemize}
\end{theorem}

\begin{remark}
If (M2) is not true, say, $\phi_l\in C^{s}$ with $v<s<k$,
then the order of the asymptotic variance of $\hat\mu^{(v)}(t)$ has the following form:
\begin{align*}
var\left(\hat\mu^{(v)}(t)\right)
&=n^{-1}(1+O\left(b^{(s-v)}\right)+O\left(b^{-2v}(T_{max}b)^{(2d-1)q}\right).
\end{align*}
\end{remark}

\begin{corollary}
Suppose that the conditions of Theorem \ref{Theorem - Bias and Var of mean and its der} hold.
For $v=0$ and $k=2$, we have 
\begin{align*}
E\left[\hat\mu(t)\right]-\mu(t)
&= b^2C_{bias,0,2}(t)(1+o(1))
\end{align*}
and
\begin{align*}
var\left(\hat\mu(t)\right)
&=n^{-1}C_{bias,0}(t)(1+o(1)).
\end{align*}
\end{corollary}

\subsection{Weak convergence of $\hat\mu^{(v)}$}
Under additional assumptions on the sequence of bandwidth $b$, weak convergence of
$\hat\mu^{(v)}(t)$ in $C[0, 1]$ in the supremum norm sense can be obtained.
\begin{theorem}
\label{Theorem - Weak convergence of mean and its der}
Let $Y_{ij}$ be defined by (\ref{Equation Model Observation})-(\ref{Equation model error}).
Suppose the assumptions in Theorem \ref{Theorem - Bias and Var of mean and its der} hold. 
Let 
$$n\to \infty,\ N\to\infty,\ b\to 0,$$
such that 
\begin{equation}
\label{Conditon - b and n}
nb^{2(k-v)}\to 0
\end{equation}
and 
\begin{equation}
\label{Condition - b and beta_N, T_N}
\lim\inf b^{2(v+2)}\beta^2_{N}(T_{max}b)^{-(2d+1)q}>C
\end{equation}
for a suitable constant $C>0$.
Then
$$Z_{n, N}(t)=:\sqrt n\left(\hat\mu^{(v)}(t)-\mu^{(v)}(t)\right)
  \Rightarrow 
  \sum_{l=1}^\infty\sqrt{\lambda_l}\phi_l^{(v)}(t)\zeta_l$$
where ``$\Rightarrow$" denotes weak convergence in $C[0, 1]$ equipped with the supremum norm,
and $\zeta_l\overset{iid}\sim N(0, 1)$.
\end{theorem}

\begin{remark}
For observed data, the Hermite rank is unknown.
Bai and Taqqu (2016) argue that a rank other than one is unstable
in the sense that, when there is a slight perturbation, it typically collapses
to rank one.
Therefore, methods to perform valid inference of the rank are needed to be developed.
As far as we know, only Beran et al (2016) performed work on designing a statistical
test based on bootstrap procedures to test the Hermite rank $m=1$ against $m>1$.
\end{remark}

\begin{remark}
Let's consider the estimation of the population mean, i.e. $v=0$ 
under the assumption of equidistance 
i.e. $\beta_N=N$, $T_{max}=N$, 
and assume the Hermit rank $q$ is $1$.
If the kernel function $K$ is of order $2$, i.e. $k=2$, 
then condition (\ref{Conditon - b and n}) reduces to $nb^4\to\infty$ and
condition (\ref{Condition - b and beta_N, T_N}) reduces to $\lim\inf N^{-2d+1}b^{3-2d}>C$.
This situation coincids with Theorem 2 in Beran and Liu (2014) and will lead to a restriction on
selecting the bandwidth (also on the relationship between $n$ and $N$), 
i.e. $CN^{-(1-2d)/(3-2d)}\leq b<< n^{-1/4}$ for some $C>0$.
As pointed out by Beran and Liu (2014), 
``In order that this can be fullfilled by a sequence of bandwidths $b_N$, we need 
$n=n_N=o(N^{4(1-2d)/(3-2d)})$.'' 
This means that the number of replicated time series cannot grow too fast compared to 
$N$, especially when $d\to 1/2$.

Under $\beta_N=N$, $T_{max}=N$, $q=1$, and $k>2$, 
condition (\ref{Conditon - b and n}) reduces to $nb^{2k}\to\infty$ and 
condition (\ref{Condition - b and beta_N, T_N}) reduces to $\lim\inf N^{-2d+1}b^{3-2d}>C$.
This situation leads to a less restrictive condition compared to order 2 kernel function, 
i.e. $CN^{-(1-2d)/(3-2d)}\leq b<< n^{-1/2k}$ for some $C>0$.
In order that this condition can be fullfilled by a sequence of bandwidths $b_N$,
we similarly need $n=n_N=o(N^{2k(1-2d)/(3-2d)})$ 
which is less restrictive, if $k\to\infty$, even for $d$ close to $1/2$.

For irregularly designed time points, the corresponding requirements
on bandwidth selection can be adapted by considering the sampling
space.
\end{remark}

\begin{remark}
Now we consider the estimation of second derivative of population mean, i.e. $v=2$.
under the assumption of equidistance and Gaussianlity of the process, 
i.e. $\beta_N=N$, $T_{max}=N$, $q=1$.

If the kernel function $K$ is of order $(2, 4)$, i.e. $k=v+2$,
condition (\ref{Conditon - b and n}) implies $b<< n^{-\frac{1}{4}}$,
and condition (\ref{Condition - b and beta_N, T_N}) reduces to $\lim\inf N^{-2d+1}b^{7-2d}>C$.
These conditions result in $CN^{-(1-2d)/(7-2d)}\leq b<< n^{-1/4}$.
We need $n=n_N=o(N^{4(1-2d)/(7-2d)})$ in order the above condition to be fullfilled.
This condition is more restrictive than the condition in mean estimation
where $n=n_N=o(N^{4(1-2d)/(3-2d)})$.

If $k>2$, condition (\ref{Conditon - b and n}) reduces to $nb^{2k}\to\infty$ and
condition (\ref{Condition - b and beta_N, T_N}) reduces to $\lim\inf N^{-2d+1}b^{7-2d}>C$.
This results in a less restrictive condition compared to order 2 kernel function, 
i.e. $CN^{-(1-2d)/(7-2d)}\leq b<< n^{-1/2k}$ for some $C>0$.
Similarly we need $n=n_N=o(N^{2k(1-2d)/(7-2d)})$ 
and this is less restrictive, if $k\to\infty$, even for $d$ close to $1/2$.

For irregularly designed time points, the corresponding requirements
on bandwidth selection can be adapted by considering the sampling
space.
\end{remark}

\section{Final Remarks}
We considered estimation of population mean and its derivatives 
in repeated time series with subordinated Gaussian long-range dependent errors
with an FDA structure.
A functional limit theorem is obtained for Priestley-Chao type kernel estimator
of $\mu^{(v)}(t)$, provided that the number of curves $n$, the number of observations $N_i$ 
on each curve and the spacings, satistify some restrictions.
These restrictions provide information for selecting the bandwidth $b_N$.

Our results allow for more flexibility in modelling repeated time series.
This flexibility is needed to extract important information from 
temporal datasets available in health for example.
Moreover, one can estimate rapid change points of $\mu(t)$ based on our results,
espeically based on the second derivative.
Two more urgent questions are the covariance estimation and functional principal
component analysis in this context, which are the fundamental and powerful tools to analysis
functional data.
These will be potential future research topics.

\section{Appendix}

\begin{proof}
For simplicity of presentation we consider the case with only one basis function
$\phi(t)$.
Therefore, let 
$$Y_{ij}=\mu(t_{ij})+\xi_i\phi(t_{ij})+\epsilon_i(j).$$

1) The bias term is straightforward obtained and follows the standard situation.
In fact, 
\begin{align*}
&E\left[\hat\mu^{(v)}(t)\right]-\mu^{(v)}(t)\\
&=\frac{(-1)^v}{b^{v+1}} \frac{1}{n}\sum_{i=1}^n
\sum_{j=1}^{N_i}(t_{ij}-t_{i,j-1}) K^{(v)}\left(\frac{t_{ij}-t}{b}\right)E[Y_{ij}]
  -\mu^{(v)}(t)\\
&=\frac{(-1)^v}{b^{v+1}} \frac{1}{n}\sum_{i=1}^n
\sum_{j=1}^{N_i}(t_{ij}-t_{i,j-1}) K^{(v)}\left(\frac{t_{ij}-t}{b}\right)\mu(t_{ij})
  -\mu^{(v)}(t)\\
&=\frac{(-1)^v}{b^{v+1}}\left[\int_0^1K^{(v)}\left(\frac{x-t}{b}\right)\mu(x)dx
                              +O\left(\beta_{N}^{-1}\right)\right]
  -\mu^{(v)}(t)\\
&=\frac{(-1)^v}{b^{v+1}}\int_{-1}^1K^{(v)}(y)\mu(by+t)bdy
  -\mu^{(v)}(t)+O\left(\left(b^{v+1}\beta_{N}\right)^{-1}\right)\\
&=\frac{(-1)^v}{b^{v}}\int_{-1}^1K^{(v)}(y)
  \left[\sum_{l=0}^k\frac{(yb)^l}{l!}\mu^{(l)}(t)+o\left(b^k\right)\right]dy
  -\mu^{(v)}(t)+O\left(\left(b^{v+1}\beta_{N}\right)^{-1}\right)\\
&= b^{k-v}C_{bias,v,k}(t)+o\left(b^{k-v}\right)+O\left(\left(b^{v+1}\beta_{N}\right)^{-1}\right).
\end{align*}
where
$$C_{bias,v,k}(t)=\frac{\mu^{(k)}(t)}{k!} \int K^{(v)}(x)x^{k}dx.$$

2) Now, we consider the asymptotic variance.
$$\hat\mu^{(v)}(t)-E\left[\hat\mu^{(v)}(t)\right]
=\frac{(-1)^v}{b^{v+1}} \frac{1}{n}\sum_{i=1}^n
\sum_{j=1}^{N_i}(t_{ij}-t_{i,j-1})K^{(v)}\left(\frac{t_{ij}-t}{b}\right)
 [\xi_i\phi(t_{ij})+\epsilon_i(j)].$$
Since $\xi$ and $\epsilon$ are independent, we have
\begin{align*}
var\left(\hat\mu^{(v)}(t)\right)=n^{-2}b^{-2(v+1)}(A_{n, N}(t)+B_{n,N}(t))
\end{align*}
where 
\begin{align*}
A_{n,N}(t)=var\left(\sum_{i=1}^n\sum_{j=1}^{N_i}(t_{ij}-t_{i,j-1})K^{(v)}
\left(\frac{t_{ij}-t}{b}\right)\xi_i\phi(t_{ij})\right),
\end{align*}
\begin{align*}
B_{n,N}(t)=var\left(\sum_{i=1}^n\sum_{j=1}^{N_i}(t_{ij}-t_{i,j-1})K^{(v)}
\left(\frac{t_{ij}-t}{b}\right)\epsilon_i(j)\right).
\end{align*}

i) For the item $A_{n, N}(t)$, since $\phi\in C^{k}[0, 1]$ and $\xi_i$ are independent, we have
\begin{align*}
A_{n,N}(t)
&=var\left(\sum_{i=1}^n\sum_{j=1}^{N_{i}}(t_{ij}-t_{i,j-1})K^{(v)}
  \left(\frac{t_{ij}-t}{b}\right)\phi(t_{ij}) \xi_i\right)\\
&=n\lambda\left(\int_0^1K^{(v)}\left(\frac{x-t}{b}\right)\phi(x)dx
     +O\left(\beta_{N}^{-1}\right)\right)^2\\
&=n\lambda\left(\int_{-1}^1K^{(v)}(y)\phi(by+t)bdy
     +O\left(\beta_{N}^{-1}\right)\right)^2\\
&=n\lambda\left(b\int_{-1}^1K^{(v)}(y)
                \left[\sum_{l=0}^k\frac{(yb)^l}{l!}\phi^{(l)}(t)+o\left(b^k\right)\right]dy
     +O\left(\beta_{N}^{-1}\right)\right)^2\\
&=nb^{2(v+1)}\lambda\left(\phi^{(v)}(t)\right)^2
  \left(1+O\left(b^{k-v}\right)+O\left(\left(b^{v+1}\beta_{N}\right)^{-1}\right)\right).
\end{align*}

ii) For the item $B_{n, N}(t)$, we have
\begin{align*}
B_{n,N}(t)
&=\sum_{i=1}^nvar\left(\sum_{j=1}^{N_i}(t_{ij}-t_{i,j-1})K^{(v)}
        \left(\frac{t_{ij}-t}{b}\right)\epsilon_i(j)\right)\\
&=\sum_{i=1}^n \sum_{j=1}^{N_i}(t_{ij}-t_{i,j-1})^2
     \left(K^{(v)}\left(\frac{t_{ij}-t}{b}\right)\right)^2
     V_{i,j}\\
&+ \sum_{i=1}^n \sum_{j_1<j_2}^{N_i}(t_{ij_1}-t_{i,j_1-1})(t_{ij_2}-t_{i,j_2-1})
     K^{(v)}\left(\frac{t_{ij_1}-t}{b}\right)K^{(v)}\left(\frac{t_{ij_2}-t}{b}\right)
     V_{i,j_1, j_2}\\
&+ \sum_{i=1}^n \sum_{j_1>j_2}^{N_i}(t_{ij_1}-t_{i,j_1-1})(t_{ij_2}-t_{i,j_2-1})
     K^{(v)}\left(\frac{t_{ij_1}-t}{b}\right)K^{(v)}\left(\frac{t_{ij_2}-t}{b}\right)
     V_{i,j_1, j_2}\\
&=: B_{n,N,1}(t)+B_{n,N,2}(t)+B_{n,N,3}(t)
\end{align*}
where
\begin{align*}
V_{i,j}
=cov(\epsilon_i(j), \epsilon_i(j))
=\sum_{l=q}^\infty\frac{c_l^2(t_{ij})}{l!}
\end{align*}
\begin{align*}
V_{i,j_1, j_2}
=cov(\epsilon_i(j_1), \epsilon_i(j_2))
=\sum_{l=q}^\infty\frac{c_l(t_{ij_1})c_l(t_{ij_2})}{l!}\gamma_Z^l(T_{ij_1}-T_{ij_2})
\end{align*}
and
$$\gamma_Z(T_{ij_1}-T_{ij_2})\sim c_Z|T_{ij_1}-T_{ij_2}|^{2d-1}.$$

For $B_{n,N,1}(t)$ we have,
\begin{align*}
|B_{n,N,1}(t)|
&=\sum_{i=1}^n \sum_{j=1}^{N_i}(t_{ij}-t_{i,j-1})^2
     \left(K^{(v)}\left(\frac{t_{ij}-t}{b}\right)\right)^2
     V_{i,j}\\
&\leq C_1\sum_{i=1}^n \sum_{j=1}^{N_i}(t_{ij}-t_{i,j-1})^2
     \left(K^{(v)}\left(\frac{t_{ij}-t}{b}\right)\right)^2\\
&=b^2C_1\sum_{i=1}^n \sum_{j=1}^{N_i}\left(\frac{t_{ij}-t_{i,j-1}}{b}\right)^2
     \left(K^{(v)}\left(\frac{t_{ij}-t}{b}\right)\right)^2\\
&\leq b^2C_1\sum_{i=1}^n \sum_{j=1}^{N_i}\left(\frac{t_{ij}-t_{i,j-1}}{b}\right)\frac{1}{b\beta_N}
     \left(K^{(v)}\left(\frac{t_{ij}-t}{b}\right)\right)^2\\
&=b^2\frac{1}{b\beta_N}C_1\sum_{i=1}^n 
  \left(\int_{\frac{t_{i1}}{b}}^{\frac{t_{iN_i}}{b}}
     \left(K^{(v)}\left(x-\frac{t}{b}\right)\right)^2dx+O\left((b\beta_N)^{-1}\right)\right)\\
&=b^2\frac{n}{b\beta_N}C_1 
  \left(\int_{-1}^{1}
     \left(K^{(v)}(u)\right)^2du+O\left((b\beta_N)^{-1}\right)\right)\\
\end{align*}
where $$C_1=\frac{\sup_{l\geq q}\sup_{t\in[0, 1]}c_l^2(t)}{e}$$

For $B_{n,N,2}(t)$, notice assumption (T2)
$$-1<(2d-1)q<0,$$
we have
$$V_{i,j_1, j_2}\sim \frac{c_q^2(t)}{q!}\gamma_Z^q(T_{j_1}-T_{j_2})\sim |T_{ij_1}-T_{ij_2}|^{(2d-1)q}$$
for $j_1, j_2\in U_b(t)$ with $U_b(t)=\{k\in\mathbb N:|t-t_{ik}|\leq b\}$.
Moreover since $K(x)=0$ for $|x|>1$, we have
\begin{align*}
&B_{n,N,2}(t)\\
&=\sum_{i=1}^n \sum_{j_1<j_2}^{N_i}(t_{ij_1}-t_{i,j_1-1})(t_{ij_2}-t_{i,j_2-1})
     K^{(v)}\left(\frac{t_{ij_1}-t}{b}\right)K^{(v)}\left(\frac{t_{ij_2}-t}{b}\right)
     (T_{ij_1}-T_{ij_2})^{(2d-1)q}\\
&=b^2(T_{max}b)^{(2d-1)q}\sum_{i=1}^n \sum_{j_1\in A_{i,j_1}}^{N_i}
     K^{(v)}\left(\frac{t_{ij_1}-t}{b}\right)\frac{t_{ij_1}-t_{i,j_1-1}}{b}\\
&\qquad \qquad \qquad \qquad \qquad \times\sum_{j_2\in A_{i,j_2}}K^{(v)}\left(\frac{t_{ij_2}-t}{b}\right)
  \left(\frac{t_{ij_1}-t_{ij_2}}{b}\right)^{(2d-1)q}
  \frac{t_{ij_2}-t_{i,j_2-1}}{b}
\end{align*}
where
$$A_{i,j_1}=\{j_1\in \mathbb N:|T_{ij_1}-tT_{max}|\leq bT_{max}\}$$
and 
$$A_{i,j_2}=\{j_2\in \mathbb N: 1\leq j_2\leq j_1-1, |T_{ij_2}-tT_{max}|\leq bT_{max}\}.$$
Setting
$$h_{N_i}(x)=K^{(v)}\left(x-\frac tb\right)\left(\frac{t_{ij_1}}{b}-x\right)^{(2d-1)q}$$
we have
\begin{align*}
&\sum_{j_2\in A_{i,j_2}}K^{(v)}\left(\frac{t_{ij_2}-t}{b}\right)
  \left(\frac{t_{ij_1}-t_{ij_2}}{b}\right)^{(2d-1)q}
  \frac{t_{ij_2}-t_{i,j_2-1}}{b}\\
&=\int_{\frac{t_{i1}}{b}}^{\frac{t_{ij_1-1}}{b}}h_{N_i}(x)dx
+\sum_{j_2\in A_{i,j_2}}h_{N_i}^\prime(x_{j_2})\left(\frac{t_{ij_2}-t_{i,j_2-1}}{b}\right)^2\\
&=\int_{\frac{t_{i1}}{b}}^{\frac{t_{ij_1-1}}{b}}h_{N_i}(x)dx
+r_{N_i,j_1}
\end{align*}
where $\frac{t_{i,j_2-1}}{b}\leq x_{ij_2}\leq\frac{t_{ij_2}}{b}$ and
$h_{N_i}^\prime(x)=g_{N_i,1}(x)+g_{N_i,2}(x)$ with
$$g_{N_i,1}(x)=K^{(v+1)}\left(x-\frac tb\right)\left(\frac{t_{ij_1}}{b}-x\right)^{(2d-1)q}$$
and
$$g_{N_i,2}(x)=K^{(v)}\left(x-\frac tb\right)\left(\frac{t_{ij_1}}{b}-x\right)^{(2d-1)q-1}(2d-1)q.$$
Notice that  
\begin{align*}
&\text{(T2): }-1<(2d-1)q<0,\\
&\text{(T4): }\alpha_{N}^{-1}\leq \underset{i}\min\{t_{ij}-t_{i,j-1}\}
\leq \underset{i}\max\{t_{ij}-t_{i,j-1}\}\leq \beta_{N}^{-1},\\
&\text{(T5): }\underset{{N}\to\infty}\lim(b\alpha_{N})^{1+(1-2d)q}(b\beta_{N})^{-2}=0, \\
&\text{(K6): }0\leq \kappa_{v+1}=\sup_{x\in[-1, 1]}|K^{(v+1)}(x)|<\infty,\\
&\text{moreover}\\
&b\beta_N\to\infty\text{ which implies }b\alpha_N\to\infty,
\end{align*} 
using the notation $k_1=[\alpha_N(t-b)]$ and $k_2=[\alpha_N(t+b)]$,
an upper bound can be given by
\begin{align*}
&\left|\sum_{j_2\in A_{i,j_2}}g_{N_i,1}(x_{j_2})\left(\frac{t_{ij_2}-t_{i,j_2-1}}{b}\right)^2\right|\\
&=\left|\sum_{j_2\in A_{i,j_2}}
K^{(v+1)}\left(x_{ij_2}-\frac tb\right)\left(\frac{t_{ij_1}}{b}-x_{ij_2}\right)^{(2d-1)q}
\left(\frac{t_{ij_2}-t_{i,j_2-1}}{b}\right)^2\right|\\
&\leq\kappa_{v+1}(b\beta_N)^{-2}
\left|\sum_{j_2=k_1}^{k_2}\left(\frac{t_{ij_1}}{b}-\frac{t_{ij_2}}{b}\right)^{(2d-1)q}\right|\\
&\leq\kappa_{v+1}(b\beta_N)^{-2}
\left|\sum_{j_2=1}^{[2b\alpha_N]}\left(\frac{j}{b\alpha_N}\right)^{(2d-1)q}\right|\\
&\leq\kappa_{v+1}b\alpha_N(b\beta_N)^{-2}\int_0^2x^{(2d-1)q}dx.
\end{align*} 
Notice that $(2d-1)q>-1$, $b\alpha_N\to\infty$, and 
$lim_{N\to\infty}(b\alpha_N)^{1+(2d-1)q}(b\beta_N)^{-2}=0$
imply $lim_{N\to\infty}(b\alpha_N)(b\beta_N)^{-2}=0$.
Moreover, $\int_0^2x^{(2d-1)q}dx<\infty$ since $(2d-1)q>-1$.
Thus, there is a uniform upper bound on $r_{N_i, j_1}$.

Analogous arguments apply to $B_{n,N,3}(t)$, and at last, we obtain 
$$B_{n, N}(t)=nb^2(T_{max}b)^{(2d-1)q}I_q(t)(1+o(1))$$
where
$$I_q(t)=\frac{c_q^2(t)}{q!}c_Z^q\int\int K(x)K(y)|x-y|^{(2d-1)q}dxdy.$$

Therefore, 
\begin{align*}
var\left(\hat\mu^{(v)}(t)\right)
&=n^{-2}b^{-2(v+1)}(A_{n, N}(t)+B_{n,N}(t))\\
&=n^{-1}\lambda\left(\phi^{(v)}(t)\right)^2\left
  (1+O\left(b^{k-v}\right)+O\left(\left(b^{v+1}\beta_{N}\right)^{-1}\right)\right)\\
&\  +n^{-1}b^{-2v}(T_{max}b)^{(2d-1)q}I_q(t)(1+o(1)).
\end{align*}

\end{proof}

\begin{proof}
Condition (\ref{Conditon - b and n}) is only reguired to make the bias of order $o(n^{-1/2})$ since
$$nb^{2(k-v)}\to 0$$
together with Theorem \ref{Theorem - Bias and Var of mean and its der}, implies
$$\lim_{n\to\infty} \sqrt n \sup_{t\in[0, 1]}
  \left|E\left[\hat\mu^{(v)}(t)\right]-\mu^{(v)}(t)\right|=0.$$
Therefore, it is sufficient to consider the process
$$Z_{n, N}^0(t):=\sqrt n\left(\hat\mu^{(v)}(t)-E\left[\hat\mu^{(v)}(t)\right]\right).$$

For simplicity of presentation, as in the proof of 
Theorem \ref{Theorem - Bias and Var of mean and its der},
we consider the case with only one basis function $\phi(t)$.
Therefore,  
$$Y_{ij}=\mu(t_j)+\xi_i\phi(t_j)+\epsilon_i(j).$$

Denote
$$c_{iN}(t)
 =\frac{(-1)^v}{b^{v+1}}\sum_{j=1}^{N_i}(t_{ij}-t_{i,j-1})
  K^{(v)}\left(\frac{t_{ij}-t}{b}\right)\phi(t_{ij}),$$
$$u_n=n^{-1/2}\lambda^{-1/2}\sum_{i=1}^n\xi_i$$
and
$$e_n(j)=n^{-1/2}\sum_{i=1}^n\epsilon_i(j),$$
we can write
$$Z_{n, N}^0(t)=S_{n,N,1}(t)+S_{n,N,2}(t)$$
where
$$S_{n,N,1}(t)=\sqrt\lambda n^{-1/2}\lambda^{-1/2}\sum_{i=1}^n\xi_ic_{iN}$$
and
$$S_{n,N,2}(t)
  =\frac{(-1)^v}{b^{v+1}}n^{-1/2}\sum_{i=1}^n
     \sum_{j=1}^{N}(t_{ij}-t_{ij-1})K^{(v)}\left(\frac{t_{ij}-t}{b}\right)\epsilon_i(j)$$
are independent.

Now, we consider $S_{n,N,1}(t)$.
Notice that, since $\xi_i\overset{iid}\sim N(0, \lambda)$, we have $u_n\sim N(0, 1)$ for all $n$.
For $c_N(t)$, we have
\begin{align*}
\left|c_{iN}(s)-c_{iN}(t)\right|
 \leq\left|\phi^{(v)}(s)-\phi^{(v)}(t)\right|
   +Cb^{k-v}+C^\star\left(b^{v+1}\beta_{N}\right)^{-1},
\end{align*}
where $C$ and $C^\star$ are suitable constants.
Thus
\begin{align*}
\omega_{S_{n,N,1}}(\Delta)
&\leq\sqrt\lambda\left(\omega_{\phi^{(v)}}(\Delta)
      +Cb^{k-v}+C^\star\left(b^{v+1}\beta_{N}\right)^{-1}\right)|u_n|,
\end{align*}
where 
$$\omega_f(\Delta)=\sup_{|s-t|\leq\Delta}|f(t)-f(s)|$$
is the modulus of continuity of function $f(t)$.
Let $\tau>0$ and $\Delta>0$ be small.
Then, we have
$$P(\omega_{S_{n,N,1}}(\Delta)>\tau)
   \leq P\left(\sqrt\lambda\left(\omega_{\phi^{(v)}}(\Delta)
        +Cb^{k-v}+C^\star\left(b^{v+1}\beta_{N}\right)^{-1}\right)|u_n|>\tau\right).$$
Taking the $\lim\sup$ over $n$ and $N$ such that conditions of 
Theorem \ref{Theorem - Weak convergence of mean and its der} hold, 
we have
$$\lim\sup P(\omega_{S_{n,N,1}}(\Delta)>\tau)
   \leq 2\left[1-\Phi\left(\tau\omega_{\phi^{(v)}}^{-1}(\Delta)\lambda^{-1/2}\right)\right],$$
where $\Phi$ denotes the cumulative distribution function of standard normal random variable.
Since $\phi^{(v)}$ is uniformly continuous on $[0, 1]$, we have
$$\lim_{\Delta\to0}\lim\sup P(\omega_{S_{n,N,1}}(\Delta)>\tau)=0.$$
Therefore, 
$$S_{n,N,1}(t)\Rightarrow \sqrt\lambda\phi^{(v)}(t)\zeta.$$

For $S_{n,N,2}(t)$,
first, we show convergence of finite-dimensional distributions.
Notice that
\begin{align*}
var(S_{n,N,2}(t))
&=b^{-2v}(T_{max}b)^{(2d-1)q}I_q(t)+r_N(t)
\end{align*}
where
$$\lim_{T_{max}\to\infty}b^{2v}(T_{max}b)^{(1-2d)q}\sup_{t\in[0,1]}|r_N(t)|=0.$$
Thus,
$$\sup_{t\in[0,1]}var(S_{n,N,2}(t))\leq C b^{-2v}(T_{max}b)^{(2d-1)q}\ (n\geq n_0,\ N\geq N_0)$$
for $n_0$ and $N_0$ large enough.
Thus, for all $p\in \mathbb N, \ t_1,...,t_p\in[0, 1]$, 
$$(S_{n,N,2}(t_1),...,S_{n,N,2}(t_p))^T\overset{d,\ p}\longrightarrow(0,...,0)^T.$$
Now we have to show the tightness of $S_{n,N,2}(t)$.
Therefore,
\begin{align*}
&E\left[(S_{n,N,2}(t)-S_{n,N,2}(s))^2\right]\\
&=\frac{1}{b^{2(v+1)}n}E[(\sum_{i=1}^n\sum_{j=1}^{N_i}(t_{ij}-t_{i,j-1})
  K^{(v)}\left(\frac{t_{ij}-t}{b}\right)\epsilon_i(j)\\
& \qquad
  -\sum_{i=1}^n\sum_{j=1}^{N_i}(t_{ij}-t_{i,j-1})
  K^{(v)}\left(\frac{t_{ij}-s}{b}\right)\epsilon_i(j))^2]\\
&\leq b^{-2(v+1)}\kappa_{v+1}^2\left(\frac{t-s}{b}\right)^2\beta_{N}^{-2}
  n^{-1}\sum_{i=1}^n \sum_{j_1,j_2=1}^{2T_{max}b} V_{i,j_1,j_2}.
\end{align*}
where
\begin{align*}
V_{i,j_1,j_2}
=Cov(\epsilon_i(j_1), \epsilon_i(j_2)
=\sum_{l=q}^\infty\frac{c_l(t_{ij_1})c_l(t_{ij_2})}{l!}\gamma_Z^l(T_{ij_1}-T_{ij_2}).
\end{align*}
Recalling that 
$$\gamma_Z(T_{ij_1}-T_{ij_2})\sim c_Z|T_{ij_1}-T_{ij_2}|^{2d-1}$$
and the condition (T2), i.e. $-1<(2d-1)q<0$, we have
$$V_{i, j_1, j_2}\sim \frac{c_q^2(t)}{q!}\gamma_Z^q(T_{ij_1}-T_{ij_2}).$$
Then, notice that $|T_{ij_1}-T_{ij_2}|\geq 1$ and assumption (T1), we have
\begin{align*}
&E\left[(S_{n,N,2}(t)-S_{n,N,2}(s))^2\right]\\
&\leq b^{-2(v+1)}\kappa_{v+1}^2(t-s)^2b^{-2}\beta_{N}^{-2}
   n^{-1}\sum_{i=1}^n \sum_{j_1,j_2=1}^{2T_{max}b}
       C_V \left|T_{ij_1}-T_{ij_2}\right|^{(2d-1)q}\\
&\leq Cb^{-2(v+1)}b^{-2}\beta_{N}^{-2}(T_{max}b)^{(2d+1)q}(t-s)^2
\end{align*}
where
$$\kappa_{v+1}=\sup_{t\in[-1, 1]}\left|K^{(v+1)}(x)\right|.$$
Therefore, assumption (\ref{Condition - b and beta_N, T_N}) $S_{n,N2}\to 0$ implies
that there is a finite constant $C^*$ such that
\begin{align*}
E\left[(S_{n,N,2}(t)-S_{n,N,2}(s))^2\right]\leq C(t-s)^2
\end{align*}
Therefore, tightness of $S_{n,N2}(t)$ ($t\in[0,1]$) and the weak convergence of
$S_{n,N,2}(t)$ to 0 in the Skorohod topology can be obtained from Billingsley (1999).

Therefore 
$$Z^0_{n,N}(t)\Rightarrow\sqrt\lambda\phi^{(v)}(t)\zeta$$
and further
$$Z_{n,N}(t)\Rightarrow\sqrt\lambda\phi^{(v)}(t)\zeta$$

The above proof can be extended to the general case with an arbitrary number of 
basis functions $\phi_l$.
\end{proof}


\begin{thebibliography}{9}
\bibitem {BaiT16} Bai, S. and Taqqu, M. S. (2016). 
How the instability of ranks in non-central limit theorems affects large-sample 
inference under long memory. 
arXiv preprint arXiv:1610.00690.

\bibitem {Beran94}J. Beran (1994).
{\it Statistics for long-memory processes},
Chapman \& Hall/CRC Press.

\bibitem {BeranF02}Beran, J. and Feng, Y. (2002).   
Local polynomial fitting with long memory, short memory and antipersistent errors.
{\it Annals of the Institute of Statisitcal Mathematics}, Vol. 54, No. 2, 291-311. 

\bibitem {Beran13}Beran, J., Feng, Y., Ghosh, S., and Kulik, R. (2013). 
{\it Long-Memory Processes: Probabilistic Properties and Statistical Methods}, 
Springer.

\bibitem {BeranL14}Beran, J. and Liu, H. (2014).   
On estimation of mean and covariance functions in repeated time series
with long-memory errors.
{\it Lithuanian Mathematical Journal}, Vol. 54, No. 1, 8-34. 

\bibitem {BeranMG16}Beran, J., M\"{o}hrle, S., and Ghosh, S. (2016). 
Testing for Hermite rank in Gaussian subordination processes. 
Journal of Computational and Graphical Statistics, 25(3), 917-934.


\bibitem {BeranW11} Beran, J. and Weiershauser, A. (2011). 
On spline regression under Gaussian subordination with long memory. 
Journal of Multivariate Analysis, Vol. 102, No. 2, 315-335.

\bibitem {Billingsley13} Billingsley, P. (2013). 
Convergence of probability measures. John Wiley and Sons.

\bibitem {BornasNB13}Bornas, X., Noguera, M.,Balle, M., Morillas-Romero, A.,  
Aguayo-Siquier, B., Tortella-Feliu, M., and Llabr\'{e}s, J. (2013).
Long-range temporal correlations in resting EEG: Its associations with 
depression-related emotion regulation strategies. 
{\it Journal of Psychophysiology}, Vol 27, No.2, 60-66.

\bibitem {CaiY10} Cai, T.T. and Yuan, M. (2010). 
Nonparametric covariance function estimation for functional and longitudinal data. 
University of Pennsylvania and Georgia inistitute of technology.

\bibitem {CsorgoM}Cs\"{o}rg\"{o}, S. and Mielniczuk, J. (1995). Nonparametric
regression under long-range dependent normal errors. 
{\it The Annals of Statistics},
Vol. 23, No. 3, 1000-1014. 

\bibitem {Gasser} Gasser, T., Muller, H.G. and Mammitzsch, V. (1985). 
Kernels for nonparametric curve estimation.
{\it Journal of the Royal Statistical Society:  Series B},
47(2), 238-252.

\bibitem {Ghosh14} Ghosh, S. (2014). On local slope estimation in partial linear models 
under Gaussian subordination. 
Journal of Statistical Planning and Inference, Vol, 155, 42-53.

\bibitem {GiraitisKS12}Giraitis, L., Koul, H.L., and Surgailis D. (2012). 
{\it Large sample inference for long memory processes}, 
Imperial College Press.


\bibitem {HorvathK12}Horv\'{a}th, L. and Kokoszka, P. (2012).
{\it Inference for Functional Data with Applications}.
Springer.

\bibitem {Linkenkaer01}Linkenkaer-Hansen, K., Nikouline, V.V., Palva, J.M., and Ilmoniemi, R.J. (2001).
Long-Range Temporal Correlations and Scaling Behavior in Human
Brain Oscillations, 
{\it The Journal of Neuroscience}, Vol. 21, No. 4, 1370-1377 (2001).

\bibitem {MenendezGB10}Men\'endez, P., Ghosh, S., and Beran, J. (2010). 
On rapid change points under long memory. 
{\it Journal of statistical planning and inference}, Vol. 140, No. 11, 3343-3354.

\bibitem {MenendezGKT13}Men\'endez, P., Ghosh, S., Kuensch, H. R., and Tinner, W. (2013). 
On trend estimation under monotone Gaussian subordination with long-memory: 
application to fossil pollen series. 
Journal of Nonparametric Statistics, Vol. 25, No. 4, 765-785.

\bibitem {Nikulin05}Nikulin, V.V. and Brismar, T. (2005).
Long-range temporal correlations in electroencephalographic oscillations: 
Relation to topography, frequency band, age and gender, 
{\it Neuroscience}, Vol. 130, No 2, 549-558.

\bibitem {PengP09}Peng, J. and Paul, D. (2009). 
A geometric approach to maximum likelihood estimation of the functional principal components 
from sparse longitudinal data. 
Journal of Computational and Graphical Statistics, 18(4), 995-1015.

\bibitem {RamsayS05}Ramsay, J.O. and Silverman, B.W. (2005).
{\it Functional Data Analysis (Second Edition)}.
Springer.

\bibitem {StaniswalisL98} Staniswalis, J.G. and Lee, J.J. (1998). 
Nonparametric regression analysis of longitudinal data. 
Journal of the American Statistical Association, Vol. 93, No 444, 1403-1418.

\bibitem {Watters00}Watters, P.A. (2000).
 Time-invariant long-range correlations in electroencephalogram dynamics. 
{\it International Journal of Systems Science}, 
Vol. 31, No. 7, 819-825.

\bibitem {YaoMW05} Yao, F., M\"uller, H. G. and Wang, J. L. (2005). 
Functional data analysis for sparse longitudinal data. 
Journal of the American Statistical Association, 100(470), 577-590.

\end{thebibliography}
\end{document}